\documentclass[preprint,12pt,authoryear]{elsarticle}
\usepackage{amssymb}
\usepackage{amsmath}
\usepackage{url}
\usepackage{cite}
\usepackage{amssymb}
\usepackage{cite}
\usepackage{amsmath,amssymb,amsfonts}
\usepackage{algorithmic}
\usepackage{graphicx}
\usepackage{textcomp}
\usepackage{xcolor}
\usepackage{tabularx}
\usepackage{tikz}
\usetikzlibrary{positioning}
\usetikzlibrary{fit}
\usepackage{booktabs} 
\usepackage{listings}
 \usepackage{xcolor}
\usepackage{color}
\usepackage[most]{tcolorbox}
\tcbuselibrary{listingsutf8}
\usepackage{framed}
\definecolor{lightgray}{rgb}{0.6, 0.6, 0.6}
\usepackage{pifont}
\newcommand{\cmark}{\ding{51}}
\newcommand{\xmark}{\ding{55}}

\journal{XYZ}

\begin{document}

\begin{frontmatter}
\title{Applied Post Quantum Cryptography: A Practical Approach for Generating Certificates in Industrial Environments} 

\author[1]{Nino Ricchizzi}

\affiliation[1]{organization={Lucerne University of Applied Sciences and Arts},
            addressline={Werftestrasse 4}, 
            city={Luzern},
            postcode={6002}, 
            state={LU},
            country={Switzerland}}
\author[2]{Christian Schwinne}

\author[2]{Jan Pelzl}
\affiliation[2]{organization={Hamm-Lippstadt University of Applied Sciences},
            addressline={Marker Allee 76-78}, 
            city={Hamm},
            postcode={59063}, 
            state={NRW},
            country={Germany}}

\begin{abstract}
The transition to post-quantum cryptography (PQC) presents significant challenges for certificate-based identity management in industrial environments, where secure onboarding of devices relies on long-lived and interoperable credentials. This work analyzes the integration of PQC into X.509 certificate structures and compares existing tool support for classical, hybrid, composite, and chameleon certificates. A gap is identified in available open-source solutions, particularly for the generation and validation of hybrid and composite certificates via command-line interfaces. To address this, a proof-of-concept implementation based on the Bouncy Castle library is developed. The tool supports the creation of classical, hybrid (Catalyst), composite, and partially chameleon certificates using PQC algorithms such as ML-DSA and SLH-DSA. It demonstrates compatibility with standard X.509 workflows and aims to support headless operation and constrained platforms typical of industrial systems. The implementation is modular, publicly available, and intended to facilitate further research and testing of PQC migration strategies in practice. A comparison with OpenSSL-based solutions highlights current limitations in standardization, toolchain support, and algorithm coverage.
\end{abstract}

\end{frontmatter}

\section{Introduction}
Quantum computing poses a fundamental threat to classical cryptographic algorithms used in public key infrastructures\citep{NISTIR8105}. In industrial systems, particularly in networked plant engineering and construction, secure key management is essential for establishing machine identities. The initial integration of devices — referred to as secure onboarding — is a crucial step to prevent unauthorized access and ensure that only trusted devices are connected\citep[p.~1--2]{nist1800-36}.
Secure onboarding mostly relies on classical digital certificates\citep{10568809}. With the advent of quantum computing, quantum computing resistant algorithms need to be integrated into certificates. With current certificate structures — classical, hybrid, and composite certificates are described in \citet{10176713}, while chameleon certificates are specified in \citet{bonnell2024chameleon} — different approaches offer varying levels of compatibility and resilience against quantum-capable adversaries.

This paper analyzes the specific challenges of integrating post-quantum cryptography (PQC) into certificate-based identity management. It compares existing approaches to certificate validation across different types, evaluates available tools (including proprietary and open-source implementations), and presents a proof-of-concept (PoC) open-source solution. The PoC addresses gaps in current toolchains and demonstrates practical feasibility for PQC adoption in industrial environments. The presented tool intents to facilitate research that needs to be done in the area of applied PCQ in practical applications.

\section{Background and Problem Statement}
Quantum computers pose a severe threat to the long-term security of public key infrastructures by rendering classical algorithms such as RSA and ECC vulnerable\citep{NISTIR8105}. In industrial systems—particularly in networked plant engineering and construction certificate based identity management is critical for secure device onboarding. These environments require long-lived credentials, offline validation, and interoperability, making cryptographic migration particularly challenging\citep{962346}.


The integration of post-quantum cryptography (PQC) into industrial key management involves multiple challenges. This paper focuses on the following representative issues:  
%
\begin{enumerate}
    \item lack of standardized support for different certificates in PKI toolchains\footnote{\url{https://pkic.org/pqccm/}}, 
    \item inconsistent validation logic across implementations\citep{wolfsslPQC}, 
    \item comparatively long deployment duration of devices in the field \citep{Hahn2016}
    \item limited compatibility with constrained or legacy devices \citep[p.~89]{Hahn2016}, and  
    \item the absence of reference implementations tailored to industrial requirements \citep[p. 100]{tno2024pqc}.
    \item migration scenarios to PQC in existing systems to provide long-term security \citep{bsi2020pqc}
\end{enumerate}
This work addresses the question: \textit{How can PQC be integrated into certificate-based identity management for secure onboarding in industrial systems using hybrid and composite certificates?}
The contribution includes a functional comparison of existing tools, identification of implementation gaps, and a proof-of-concept open source solution.

\section{Related Work}

Various approaches for integrating post-quantum cryptography (PQC) into certificate infrastructures have been proposed.

These include quantum-safe certificates with single PQC algorithms, as well as hybrid, composite,  parallel certificate structures shon in \citet{10176713}. Chameleon certificates are specified in \citet{bonnell2024chameleon}

\begin{figure}
    \centering
    \includegraphics[width=1\linewidth]{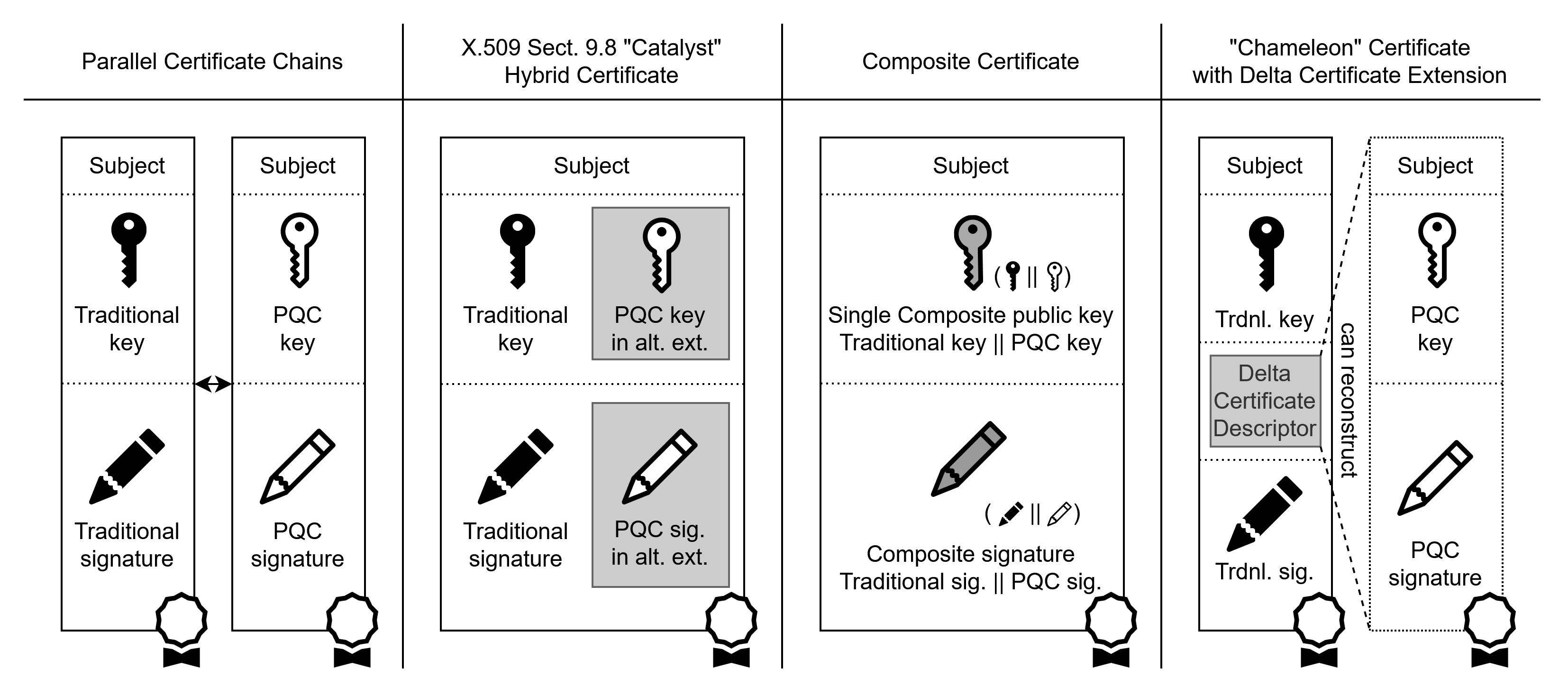}
    \caption{Overview of different hybrid PQC certificate approaches}
    \label{fig:hybrid_cert_overview}
\end{figure}

The following list provides a brief overview of the most relevant approaches. Figure \ref{fig:hybrid_cert_overview} illustrates the structural differences between these certificate types.

\begin{itemize}
  \item \textbf{Pure PQC Certificate}:  
  Contains only a post-quantum public key and a PQC signature. No fallback to classical cryptography.

  \item \textbf{Parallel Certificate Chains}:  
  Classical and PQC certificates are issued and validated independently in parallel chains.

  \item \textbf{Hybrid Certificate (X.509 Sect. 9.8 "Catalyst")}:  
  Classical certificate structure with PQC key and signature embedded in alternative extensions.

  \item \textbf{Composite Certificate}:  
  Combines classical and PQC keys into a single composite public key and includes a composite signature consisting of both algorithms.

  \item \textbf{Chameleon Certificate}:  
  A classical certificate extended with a Delta Certificate descriptor enabling the algorithmic reconstruction of a PQC-equivalent certificate. Balances backward compatibility with PQC readiness.
\end{itemize}

While conceptually distinct, many of these approaches are not yet standardized and are subject to ongoing discussion in relevant working groups (e.g., IETF\footnote{\url{https://www.ietf.org/blog/pquip/}}, ETSI\citep{etsi2023tr103966}). 

Wang et. al provide an overview of most of these certificate types, including their advantages and challenges in terms of validation, compatibility, and deployment\citep{10176713}.

The Public Key Infrastructure Consortium\footnote{\url{https://pkic.org/pqccm/}} offers a comparative overview of current tool support for each certificate type. However, this survey reveals a notable gap: there is currently no openly available, non-proprietary toolset that supports hybrid or composite certificate creation and validation via a command-line interface.

This gap motivates the development of a reference implementation. The proof-of-concept (PoC) presented in this work addresses this need by providing an open-source, CLI-based solution for hybrid and composite certificate handling with PQC algorithms. The PoC is tailored to requirements of industrial environments, including headless operation, constrained platforms, and the need for cryptographic agility in long-lived device deployments. The implementation is available via a public Git repository as part of this contribution.

\section{Toolchain Analysis}

OpenSSL is a widely used open-source toolkit for implementing TLS and managing public key infrastructures via a command-line interface (CLI). Certificate generation follows a structured process involving key pair creation, certificate signing requests (CSRs), and certificate issuance using the \texttt{openssl req} and \texttt{openssl ca} commands. Configuration is defined via an \texttt{openssl.cnf} file, which governs fields such as subject, extensions, and object identifiers (OIDs). OIDs are globally unique identifiers used to unambiguously specify cryptographic algorithms, certificate extensions, and other protocol elements in ASN.1-based structures such as X.509. Example commands are shown in Listing \ref{fig:cliopenssl}, which illustrate typical usage patterns; however, these represent only a limited excerpt. Comprehensive documentation is available at the official OpenSSL Wiki\footnote{\url{https://wiki.openssl.org}}.

\begin{figure}[htb]
\begin{lstlisting}[language=bash,
                   commentstyle=\color{lightgray},
                   caption={Examples of the OpenSSL Syntax},
                   label={fig:cliopenssl},
                   frame=single] 
# Generate a private RSA key
openssl genpkey -algorithm RSA -out private_key.pem \
-pkeyopt rsa_keygen_bits:2048

# Create a self-signed certificate
openssl req -x509 -new -key private_key.pem \
-out certificate.pem -subj "/CN=example.com"

# Certificate verification
openssl verify -CAfile ca_cert.pem certificate.pem
\end{lstlisting}
\end{figure}

We retain the CLI-based OpenSSL workflow due to its well-established usage in certificate management and integration into existing automation processes. Our certificate generation approach follows this model to ensure that the resulting artifacts can be verified using other standard-compliant tools. Since object identifiers (OIDs) for hybrid and composite certificates are not yet fully standardized, our implementation relies on temporary OIDs defined by the Bouncy Castle (BC) framework, which mostly aligns OID use with the ones outlined in the respective drafts. This means that, particularly in regard to composite certificates, future standard-compliant implementations are likely not able to interpret the custom algorithm OID for composite algorithms \texttt{1.3.6.1.4.1.18227.2.1} used by BC v1.80. As of April 2025, no standardized OIDs for any composite algorithms have been issued by IANA yet.

The objective of our PoC is to generate and verify hybrid certificates based on dual-algorithm signature schemes, while ensuring interoperability with standard-compliant X.509 implementations. Once standard OIDs and certificate formats are finalized and integrated into mainstream OpenSSL releases, our tool will remain compatible through its modular OID mapping and extensible CSR logic.
The PoC also aims to demonstrate a simple to use CLI, to facilitate familiarization and subsequent adoption not only within the circle of PQC experts, but also PKI administrators, e.g. in industrial plants.

Neither Open Quantum Safe (including \texttt{liboqs} and \texttt{oqs-provider} for OpenSSL) nor any other open-source tool known to the authors currently provides functionality to create X.509 certificates in all of the various post-quantum formats discussed — namely classical, hybrid, composite, chameleon and parallel — according to the overview presented at \url{https://pkic.org/pqccm/}. Specifically, \texttt{oqs-provider} has no support for hybrid certificates using the \texttt{subjectAltPublicKeyInfo}, \texttt{altSignatureAlgorithm}, and \\
\texttt{altSignatureValue} certificate extensions as standardized in ITU-T Rec. X.509 (10/19) Section 9.8. 
This gap in tool support motivates the development of a proof-of-concept (PoC) implementation, as outlined in the following section.


\section{Proof of Concept (PoC)}
A method for integrating PQC algorithms is provided by Open Quantum Safe through the \texttt{liboqs} library and its Python bindings. While OpenSSL, when combined with the OQS provider, enables the issuance of PQC-based certificates, it does not support all certificate types — in particular, composite certificates — as confirmed by our findings. In this proof-of-concept (PoC), we use Bouncy Castle (BC), as it offers native support for all relevant X.509 certificate types according to \url{https://pkic.org/pqccm/}. Bouncy Castle is a cryptographic library implemented in Java and C\#, supporting a wide range of algorithms and X.509 structures.
The PoC supports the creation of classical, hybrid, chameleon and composite certificates. Certificate generation follows a structured process: a cryptographic key pair is generated (e.g., RSA, Dilithium, SPHINCS+), an \texttt{X509v3CertificateBuilder} is populated with subject, issuer, and validity period, and the certificate is signed using a \texttt{ContentSigner} configured with the selected algorithm(s). 
Algorithm identifiers (OIDs) are automatically handled by Bouncy Castle and embedded based on the selected key and signature algorithm. For PQC schemes, the library uses Bouncy Castle-specific OIDs, unless standardized identifiers are available.\\
A command-line interface (CLI) is provided in the PoC to enable reproducible certificate generation. Representative usage examples are shown in Listining \ref{fig:pqcli}.

\begin{figure}[htb]
\begin{lstlisting}[language=bash,
                   commentstyle=\color{lightgray},
                   caption={Examples of the pqcli command-line usage},
                   label={fig:pqcli},
                   frame=single]

# Generate a PQC certificate (ML-DSA)
java -jar pqcli.jar cert -newkey ML-DSA:3 -subj CN=Sol

# Generate a hybrid certificate (RSA + ML-DSA)
java -jar pqcli.jar cert -newkey RSA,ML-DSA:3

# Generate a composite certificate (ML-DSA + RSA)
java -jar pqcli.jar cert -newkey ML-DSA_RSA

# Generate a SPHINCS+ (SLH-DSA) keypair
java -jar pqcli.jar key -t slh-dsa:192f

# View an existing certificate
java -jar pqcli.jar view certificate.pem
\end{lstlisting}
\end{figure}
The CLI is modular and extensible. Commands such as \texttt{cert}, \texttt{key}, and \texttt{view} are fully implemented; others (e.g., \texttt{csr}, \texttt{verify}) are planned. The full implementation, including parameter documentation, build instructions, and additional examples, is available in the associated Git repository\footnote{{\url{https://github.com/pqcli}}}.

\section{Discussion}
Table~\ref{tab:comparison} outlines the support for various important functions for hybrid PQC certificate management. Notably, the implemented proof-of-concept supports creating standard X.509 hybrid certificates using either the Catalyst or Composite approaches, and although of limited practical use, can combine both approaches to create a hybrid catalyst composite certificate, which could be interesting for validation and further research. 

\begin{table}[h!]
\centering
\begin{tabularx}{\textwidth}{|X|X|X|X|}
\hline
\textbf{Feature} & \textbf{PQCLI PoC} & \textbf{OpenSSL 3.5} & \textbf{OpenSSL 3.3 (oqs-provider)} \\
\hline
 (Standard) PQC algorithm support & ML-DSA and SLH-DSA\footnotemark[1] & ML-KEM, ML-DSA and SLH-DSA & Various (NIST Round 3 and more) \\
\hline
Composite key generation & \cmark \, Yes & \xmark \, No & \cmark \, Certain algorithm combinations \\
\hline
Composite certificate  & \cmark \, Yes & \xmark \, No & \cmark \, Certain algorithm combinations \\
\hline
Hybrid "Catalyst" certificate  & \cmark \, Yes & \xmark \, No & \xmark \, No \\
\hline
Chameleon certificate  & \xmark \, No\footnotemark[1] & \xmark \, No & \xmark \, No \\
\hline
\end{tabularx}
\caption{Comparison of capabilities of our PoC with OpenSSL 3.5 and OpenSSL 3.3 with oqs-provider}
\label{tab:comparison}
\end{table}
\footnotetext[1]{Support for additional algorithms and Chameleon certificate generation is planned.}


\section{Future Work}
Future work includes the integration of finalized object identifiers (OIDs) for hybrid and composite certificate structures as soon as standardization progresses. Several CLI commands and certificate options, such as \texttt{csr}, \texttt{verify}, \texttt{sign}, and advanced CA handling, are not yet fully implemented. The current development status and roadmap are documented in the Git repository. Additionally, a performance comparison with OpenSSL-based toolchains is planned to evaluate runtime behavior, signature sizes, and compatibility in resource-constrained environments.

\section{Conclusion}
This work addressed the integration of post-quantum cryptography (PQC) into certificate-based identity management, with a focus on X.509 certificate structures in industrial environments. A functional comparison of existing toolchains revealed a lack of support for hybrid and composite certificates in widely used open-source solutions. To address this gap, we presented a proof-of-concept implementation based on Bouncy Castle, enabling the generation and inspection of classical, hybrid, and composite certificates using PQC algorithms via a command-line interface. While the tool provides a practical and extensible foundation, it is not intended as a replacement for established toolchains such as OpenSSL, but as a complementary reference implementation for research and industrial adoption.

\section*{Acknowledgment}
Part of the research of this paper has been done within the Trustpoint project. Trustpoint was launched in September 2023 as a joint project of the Campus Schwarzwald, small and medium-sized enterprises PrimeKey Labs GmbH / Keyfactor, asvin GmbH, achelos GmbH, and the Hamm-Lippstadt University of Applied Sciences. This cooperation is made possible by the Federal Ministry of Education and Research within the framework of the SME-innovative initiative.


\bibliographystyle{apalike}

\bibliography{CertPQC} 

\end{document}